\documentclass[preprint,aps,showpacs]{revtex4}
\usepackage{amsfonts}
\usepackage{amsmath}
\usepackage{amssymb}
\usepackage{graphicx}
\usepackage{epsfig}
\begin{document}

\title{Fake state attack on practically decoy state quantum key distribution}
\author{Yong-gang Tan}
\email{yonggang.tan@gmail.com} \affiliation{Physics and Information
Engineering Department, Luoyang Normal College, Luoyang 471022,
Henan, People's Republic of China}
\begin{abstract}

In this paper, security of practically decoy state quantum key
distribution under fake state attack is considered. If quantum key
distribution is insecure under this type of attack, decoy sources
can not also provide it with enough security. Strictly analysis
shows that Eve should eavesdrop with the aid of
photon-number-resolving instruments. In practical implementation of
decoy state quantum key distribution where statistical fluctuation
is considered, however, Eve can attack it successfully with
threshold detectors.



\end{abstract}
\maketitle

Quantum key distribution (QKD) is an important application of
quantum information, with which two distant parties (the information
sender, Alice and the information receiver, Bob) can share a string
of secure key with the presence of an eavesdropper,
Eve~\cite{Bennett84,Ekert91,Gisin02}. It has been proven that
quantum principles can provide it with unconditional security when
it is implemented with ideal devices~\cite{Mayers01,Lo99,Shor00}. In
practical implementation of QKD, however, real-life devices are
taken used. They are imperfect and apt to some sophisticated
eavesdropping~\cite{Huttner95,Brassard00,Lutkenhaus02,Qi07,Fung07,Zhao08,Makarov05,Lydersen10,Yuan10,Lydersen101,Gerhardt11},
part of which have been realized with lab settings. Furthermore, in
QKD realization, Alice and Bob's experimental conditions are assumed
to be based on present technology but Eve's ability is only limited
by quantum principles. Then those eavesdropping schemes still
lacking experimental demonstration should also be considered
seriously. Recently, fake state attack is experimentally proven to
be fatal to some commercial quantum key distribution systems, with
which the latent eavesdropper can obtain full information shared
between Alice and Bob without been
detected~\cite{Makarov05,Lydersen10,Yuan10,Lydersen101,Gerhardt11}.

The fake state attack is a type of intercept-resent attack, where
Eve blocks all Alice's pulses and measures them on randomly chosen
bases. Then she prepares her measurement results on fresh pulses and
transfers them to Bob. At the same time, she controls Bob's
detectors to work in linear mode: if Bob has the same basis choices
as Eve, Eve's pulses will provide enough power above the threshold
value to generate triggers and Bob gets Eve's bit values; when their
basis choices are different, however, Eve's pulses are split below
the threshold intensity and unable to introduce click on Bob's
detectors. It is apparent that Eve's intervention introduces
tolerably error rate and generates identical key string as the legal
users'. Then Alice and Bob will acknowledge the validity of their
key and ignore Eve's presence. With a half probability, Bob and Eve
will have the same choices on their measurement bases. Thus Eve may
eavesdrop on their communication successfully if the combining
efficiency of the quantum channel and the measurement devices is
less than $\frac{1}{2}$.

Suppose Alice has a weaken coherent source whose photon number obeys
Poisson distribution
\begin{equation}
p_{n}(\mu)=\frac{\mu^{n}}{n!}e^{-\mu},
\end{equation}
where $\mu$ is average intensity of the source and $n$ is photon
number of the incoming pulses. She randomly chooses her bases and
prepares her bit values on the pulses, then she transfers them to
Bob. To avoid being caught on line, Eve must make Bob's gains and
error rates identical to that when there is no eavesdropper. That
is, Eve must make her eavesdropping satisfy
\begin{equation}
\begin{array}{lll}
\frac{1}{2}\eta_{f}\sum_{n=1}^{\infty}p_{n}(\mu)&=&\sum_{n=0}^{\infty}p_{n}(\mu)[1-(1-p_{d})^{2}(1-\eta)^{n}],\\
e_{f}\eta_{f}\sum_{n=1}^{\infty}p_{n}(\mu)&=&\sum_{n=0}^{\infty}p_{n}(\mu)\{1-(1-p_{d})^{2}(1-\eta)^{n}-(1-p_{d})\\
&\times&[(1-{\eta}e_{d})^{n}-(1-\eta+{\eta}e_{d})^{n}]\}.
\end{array}
\end{equation}
Here $\eta_{f}$ is the probability Eve will prepare fresh pulses
according to her measurement results, $e_{f}$ is the probability she
prepares wrong bit value on the fresh pulses. And $e_{d}$ is the
probability of misalignment between Alice and Bob. If Eve can keep
alignment between her measurement bases and Alice's preparing bases,
and that between her preparing bases and Bob's measurement bases at
the same time, she can control the error rate on Bob's results at
her will. Furthermore, practical QKD system is very
lossy~\cite{Gobby04}, the relationships in Eq. (2) should be
simulated by Eve easily. Thus if Eve can blind Bob's detectors, the
QKD system will be totally insecure.

In practical implementation of QKD, decoy sources are usually added
in~\cite{Hwang03,Lo05,Wang05}. They have the same characters with
that of signal source apart from their average intensities, that is
\begin{equation}
p_{n}(\nu)=\frac{\nu^{n}}{n!}e^{-\nu}.
\end{equation}
Alice randomly encodes her bit value on signal source or decoy
sources. Eve can not tell the decoy sources from the signal source,
then she must treat all sources in the same way. Furthermore, she
should mock the loss and noise in the quantum channel. Or else, her
intervention will inevitably introduce different disturbances on
Bob's results from different sources. It is easy to verify that the
relationship in Eq. (2) can not be met for the signal source and the
decoy sources at the same time. In order to eavesdrop on the decoy
state QKD, Eve should have the ability of differentiating photon
number, with which she can treat pulses with the same photons
similarly.

In photon-number-splitting (PNS) attack, Eve is assumed to have
ability of differentiating photon number in pulses without affect
their polarizations~\cite{Huttner95,Brassard00,Lutkenhaus02}. Then
quantum nondemolition (QND) measurement on photon number is
required, and this is still missed with present technology. In the
fake state attack, however, Eve can measure polarizations on the
pulses directly, which means she can get photon number and
polarization of the pulses at the same time with
photon-number-resolving detectors~\cite{Rosenberg05,Hadfield09}. If
she has the ability of differentiate photon number, she can treat
pulses with the same photons in a similar way. In order to simulate
the lossy and noise for all sources, Eve's eavesdropping should
satisfy
\begin{equation}
\begin{array}{l}
\frac{1}{2}\eta_{n}^{f}=1-(1-p_{d})^{2}(1-\eta)^{n},\\
e_{n}^{f}\eta_{n}^{f}=1-(1-p_{d})^{2}(1-\eta)^{n}-(1-p_{d})[(1-{\eta}e_{d})^{n}-(1-\eta+{\eta}e_{d})^{n}].
\end{array}
\end{equation}
Here $\eta_{n}^{f}$ and $e_{n}^{f}$ have similar definition as that
in Eq. (2), but they corresponds to pulses with definite photon
number $n$. Eve detects nothing from Alice when $n=0$, however, she
should prepare random-bit-value pulses with probability
$4p_{d}-2p_{d}^{2}$ in order to simulate dark count rate on Bob's
detectors. Noticing that Eq. (4) has nothing to do with $p_{n}(\mu)$
and $p_{n}(\nu)$, thus Eve can eavesdrop on the decoy state QKD
successfully if she has a set of photon-number-resolving detectors.

Now it is interesting whether Eve can eavesdrop on the decoy state
QKD protocol without photon-number-resolving detectors. However, as
mentioned before, eavesdropping on the decoy sources should have
similar relationships as those in Eq. (2), that is,
\begin{equation}
\begin{array}{lll}
\frac{1}{2}\eta_{f}\sum_{n=1}^{\infty}p_{n}(\nu)&=&\sum_{n=0}^{\infty}p_{n}(\nu)[1-(1-p_{d})^{2}(1-\eta)^{n}],\\
e_{f}\eta_{f}\sum_{n=1}^{\infty}p_{n}(\nu)&=&\sum_{n=0}^{\infty}p_{n}(\nu)\{1-(1-p_{d})^{2}(1-\eta)^{n}-(1-p_{d})\\
&\times&[(1-{\eta}e_{d})^{n}-(1-\eta+{\eta}e_{d})^{n}]\}.
\end{array}
\end{equation}
And one can find that there are not such $\eta_{f}$ and $e_{f}$ to
meat the relationships in Eq. (2) and those in Eq. (5) at the same
time. However, considering the imperfection of practical
implementation of decoy state QKD, the relationships in Eq. (2) and
Eq. (5) may be loosen.

Experimentally, Alice and Bob's key is distributed within a finite
period of time, generally in several hours~\cite{Ma05}. Then the
pulses generated by Alice should also be finite. We assume the
number of pulses emitted from the sources is $N=10^{10}$ in the
following discussion. If Bob's expectingly detections is $p_{det}$
under ideal circumstance, and his detecting events with finite
resources is $p_{det}^{\prime}$, $p_{det}^{\prime}$ should deviate
from $p_{det}$ with a small fluctuation $\delta_{p_{det}}$. The
probability $P(|p_{det}-p_{det}^{\prime}|>\delta_{p_{det}})$ can be
estimated to be less than
$q=\exp(-\frac{N\delta_{p_{det}}^{2}}{4p_{det}})$. If we require
$q=\exp(-25)$, $\frac{\delta_{p_{det}}^{2}}{p_{det}}$ can be
estimated to be $10^{-8}$ approximately, and $\delta_{p_{det}}$ can
be calculated as $10^{-4}p_{det}^{\frac{1}{2}}$ accordingly. Thus in
practical implementation of fake state attack, Eve should ensure
Bob's detecting events satisfy
\begin{equation}
p_{det}-10^{-4}p_{det}^{\frac{1}{2}}<p_{det}^{\prime}<p_{det}+10^{-4}p_{det}^{\frac{1}{2}}.
\end{equation}

Strictly speaking, the statistical fluctuations on different sources
are usually not the same as the number of pulses generated from
different sources may be not assigned to be the same. At the same
time, the probabilities of detecting events for different sources
are not identical because of their disparate intensities. As the
total number of the pulses is $N$, the practical number of pulses
assigned for different sources should be less than this value, the
practically tolerant fluctuation should be greater than that in Eq.
(6). It means Alice and Bob should accept the validity of their
results if the statistical fluctuations on the detecting events from
every sources satisfy the relationship in Eq. (6). And similar
relationship can be obtained for the gain of QBER on every source,
that is
\begin{equation}
p_{err}-10^{-4}p_{err}^{\frac{1}{2}}<p_{err}^{\prime}<p_{err}+10^{-4}p_{err}^{\frac{1}{2}}.
\end{equation}

It is apparent that Bob's expecting detecting results should be
$p_{det}(\mu)=1-(1-p_{d})^{2}e^{-\eta\mu}$ for signal source, and
$p_{det}(\nu)=1-(1-p_{d})^{2}e^{-\eta\nu}$ for decoy sources. And
the actual detecting events for them should be
$p_{det}^{\prime}(\mu)=\frac{1}{2}\eta_{f}(1-e^{-\mu})$ and
$p_{det}^{\prime}(\nu)=\frac{1}{2}\eta_{f}(1-e^{-\nu})$
respectively. Similarly, the expecting error rate for signal source
and decoy sources are
$p_{err}(\mu)=\frac{1}{2}\sum_{n=0}^{\infty}p_{n}(\mu)\{1-(1-p_{d})^{2}(1-\eta)^{n}-(1-p_{d})[(1-{\eta}e_{d})^{n}-(1-\eta+{\eta}e_{d})^{n}]\}$
and
$p_{err}(\nu)=\frac{1}{2}\sum_{n=0}^{\infty}p_{n}(\nu)\{1-(1-p_{d})^{2}(1-\eta)^{n}-(1-p_{d})[(1-{\eta}e_{d})^{n}-(1-\eta+{\eta}e_{d})^{n}]\}$.
And the actual error rate for them can be calculated as
$p^{\prime}_{err}(\mu)=\frac{1}{2}e_{f}\eta_{f}(1-e^{-\mu})=e_{f}p^{\prime}_{det}(\mu)$
and
$p^{\prime}_{err}(\nu)=\frac{1}{2}e_{f}\eta_{f}(1-e^{-\nu})=e_{f}p^{\prime}_{det}(\nu)$.

We can estimate the feasibility of Eve's attack with the
experimental parameters in~\cite{Gobby04}, that is,
$p_{d}=8.5\times10^{-7}$, $\eta_{B}=4.5\%$, $e_{d}=3.3\%$ and loss
coefficient $\alpha$ in the quantum channel is $0.21$ dB/km. If the
transmission distance between Alice and Bob is $120$km, one can
obtain $\eta=1.359\times10^{-4}$. When there are only two sources,
that is, a signal source with intensity $\mu=0.479$ and a weaker
decoy state with intensity $\nu=0.127$~\cite{Ma05}. As the
statistical fluctuations on the the results of signal source satisfy
the relationships in Eq. (6) and Eq. (7), its $\eta_{f}$ should
range from $3.467\times10^{-4}$ to $3.553\times10^{-4}$, and its
$e_{f}$ can range from $4.178\times10^{-2}$ to $4.806\times10^{-2}$.
Similarly, statistical fluctuation on the the results of weaker
decoy source should satisfy the relationships in Eq. (6) and Eq.
(7), its $\eta_{f}$ can be calculated to range from
$3.106\times10^{-4}$ to $3.252\times10^{-4}$, and its $e_{f}$ ranges
from $6.705\times10^{-2}$ to $8.307\times10^{-2}$. As there is no
overlap on the parameters of both sources, it seems that Eve can not
eavesdrop on the decoy state QKD protocol with threshold detectors.

Noticing that the dark count rate functions importantly in practical
implementation of decoy state QKD protocol when the transmission
distance is comparably long. Furthermore, though threshold detectors
can not tell the photon number in the incoming pulses, they can
differentiate vacuum pulses from non-vacuum pulses. Then it may help
Eve with her eavesdropping if she treats the vacuum pulses and
non-vacuum pulses in different ways. She prepares random-bit-value
pulses with probability $4p_{d}-2p_{d}^{2}$ for Bob when she
detecting nothing from Alice. It is easily verified that Eve's
eavesdropping results on the vacuum pulses coincide well with what
Bob expecting for. When there is nonvacuum pulses, she makes fresh
pulses according to her results with probability $\eta_{f}$ and
introduces error on them with probability $e_{d}$. Here Eve makes
error on the nonvacuum pulses with probability $e_{d}$ because
errors introduced on nonvacuum pulses are mainly introduced by
misalignment between Alice and Bob. Then for signal source, one can
obtain
\begin{equation}
\begin{array}{lll}
\frac{1}{2}\eta_{f}\sum_{n=1}^{\infty}p_{n}(\mu)&=&\sum_{n=1}^{\infty}p_{n}(\mu)[1-(1-p_{d})^{2}(1-\eta)^{n}],\\
\eta_{f}e_{d}\sum_{n=1}^{\infty}p_{n}(\mu)&=&\sum_{n=1}^{\infty}p_{n}(\mu)\{1-(1-p_{d})^{2}(1-\eta)^{n}-(1-p_{d})\\
&\times&[(1-{\eta}e_{d})^{n}-(1-\eta+{\eta}e_{d})^{n}]\}.
\end{array}
\end{equation}
Similar relationship can also be obtained for decoy source.

The probability of expecting detections $p_{det}$ both for signal
source and decoy source can still be calculated as that above.
However, the actual detections $p^{\prime}_{det}$ for them are
altered slightly. That is,
$p^{\prime}_{det}(\mu)=e^{-\mu}[1-(1-p_{d})^{2}]+\frac{1}{2}\eta_{f}(1-e^{-\mu})$
and
$p^{\prime}_{det}(\nu)=e^{-\nu}[1-(1-p_{d})^{2}]+\frac{1}{2}\eta_{f}(1-e^{-\nu})$.
Similarly, the expressions for $p_{err}(\mu)$ and $p_{err}(\nu)$ are
still the same. And $p^{\prime}_{err}(\mu)$ and
$p^{\prime}_{err}(\nu)$ should be recalculated as
$\frac{1}{2}e^{-\mu}[1-(1-p_{d})^{2}]+\frac{1}{2}e_{d}\eta_{f}(1-e^{-\mu})$
and
$\frac{1}{2}e^{-\nu}[1-(1-p_{d})^{2}]+\frac{1}{2}e_{d}\eta_{f}(1-e^{-\nu})$
respectively. As their statistical fluctuations should still be
bounded with the relations in Eq. (6) and Eq. (7). With simple
calculation, we find there is no such $\eta_{f}$ for signal source,
and $\eta_{f}$ for decoy source ranges from $2.855\times10^{-4}$ to
$3.001\times10^{-4}$. That is, this scheme is still inefficient in
helping Eve to eavesdrop on decoy state QKD protocol with threshold
detectors.

Eve takes control the whole quantum channel, however, she may not
set her eavesdrop point adjacent to Alice's lab. Her intervention
site may be anywhere between Alice's and Bob's labs. We will show
that this change will help Eve to Eavesdrop on the decoy state QKD
protocol successfully with threshold detectors. Let the distance
between Alice's lab and Bob's eavesdropping site be $l$ km, it is
apparent smaller $l$ requires ability to discriminate photon number
in the pulses, and larger $l$ may lead to failure of her blinding
attack. Then the optimal site should have largest $l$ where Eve can
carry out her eavesdropping successfully. The transmission
efficiency at this point can be calculated as
$\eta_{l}=10^{-\frac{\alpha{l}}{10}}$. And the statistical
distribution in the incoming pulses can be represented as
\begin{equation}
\begin{array}{lll}
p_{n}^{l}(\mu)&=&\frac{(\eta_{l}\mu)^{n}}{n!}e^{-\eta_{l}\mu},\\
p_{n}^{l}(\nu)&=&\frac{(\eta_{l}\nu)^{n}}{n!}e^{-\eta_{l}\nu}.
\end{array}
\end{equation}

If Eve takes her eavesdropping scheme as that in Eq. (8), one can
obtain
\begin{equation}
\begin{array}{lll}
\frac{1}{2}\eta_{f}\sum_{n=1}^{\infty}p_{n}^{l}(\mu)&=&\sum_{n=1}^{\infty}p_{n}^{l}(\mu)[1-(1-p_{d})^{2}(1-\eta^{\prime})^{n}],\\
\eta_{f}e_{d}\sum_{n=1}^{\infty}p_{n}^{l}(\mu)&=&\sum_{n=1}^{\infty}p_{n}^{l}(\mu)\{1-(1-p_{d})^{2}(1-\eta^{\prime})^{n}-(1-p_{d})\\
&\times&[(1-{\eta^{\prime}}e_{d})^{n}-(1-\eta^{\prime}+{\eta^{\prime}}e_{d})^{n}]\},
\end{array}
\end{equation}
with $\eta^{\prime}=\eta_{b}10^{-\frac{120-l}{10}}$ for signal
source. And similar relationships can be obtained for decoy source
\begin{equation}
\begin{array}{lll}
\frac{1}{2}\eta_{f}\sum_{n=1}^{\infty}p_{n}^{l}(\nu)&=&\sum_{n=1}^{\infty}p_{n}^{l}(\mu)[1-(1-p_{d})^{2}(1-\eta^{\prime})^{n}],\\
\eta_{f}e_{d}\sum_{n=1}^{\infty}p_{n}^{l}(\nu)&=&\sum_{n=1}^{\infty}p_{n}^{l}(\mu)\{1-(1-p_{d})^{2}(1-\eta^{\prime})^{n}-(1-p_{d})\\
&\times&[(1-{\eta^{\prime}}e_{d})^{n}-(1-\eta^{\prime}+{\eta^{\prime}}e_{d})^{n}]\}.
\end{array}
\end{equation}
As Eve's detecting efficiency is very lower, it is easy to verify
that Eve can set $l=120$ km. We can then obtain $\eta_{f}$ ranging
from $8.893\times10^{-2}$ to $9.120\times10^{-2}$ for signal source,
and it ranges from $8.775\times10^{-2}$ to $9.229\times10^{-2}$ for
decoy source. Then Eve can launch fake state attack at $l=120$ km
with threshold detectors just by preparing what she have measured
with probability $\eta_{f}$ ranging from $8.893\times10^{-2}$ to
$9.120\times10^{-2}$, and she introduces error on them with
probability $e_{d}$.

Then when statistical fluctuation is considered, Eve can
eavesdropping on decoy state QKD even with threshold detectors. She
may fail to eavesdrop successfully when her intervention site is
closer to Alice's lab, and numerical simulation shows it may be
easier for her to attack on this protocol when her intervention site
is farer away from Alice's lab. This is because the nearer to Bob's
lab, the greater probability of single-photon pulses for nonvacuum
pulses can be obtained. Then Eve can omit the effect of multi-photon
pulses treat all pulses as single photons. In practical decoy state
QKD protocol, Alice may introduce vacuum decoy state to estimate the
dark counts on Bob's detectors.~\cite{Ma05}. As Eve prepares
random-bit-value pulses with probability $4p_{d}-2p_{d}^{2}$ when
she detects nothing, however, the statistical fluctuations on the
vacuum decoy state can be verified to be met automatically. In order
to understand Eve's eavesdropping better, we give a simulation
numerically on the relation between Eve's $\eta_{f}$ and her
intervention site $l$, as is plotted as that in Fig. 1. It shows
that there is not suitable $\eta_{f}$ and for signal source when
$l\le10$ km. Furthermore, Eve can not launch her fake state attack
on this protocol when her intervention site $l$ is less than $30$ km
as there is no overlap on $\eta_{f}$ for both sources. When $l$ is
greater than $45$ km, one can find the suitable $\eta_{f}$ for
signal source also suits for decoy source.

\begin{figure}[!ht]
\centering
\includegraphics[width=1\textwidth]{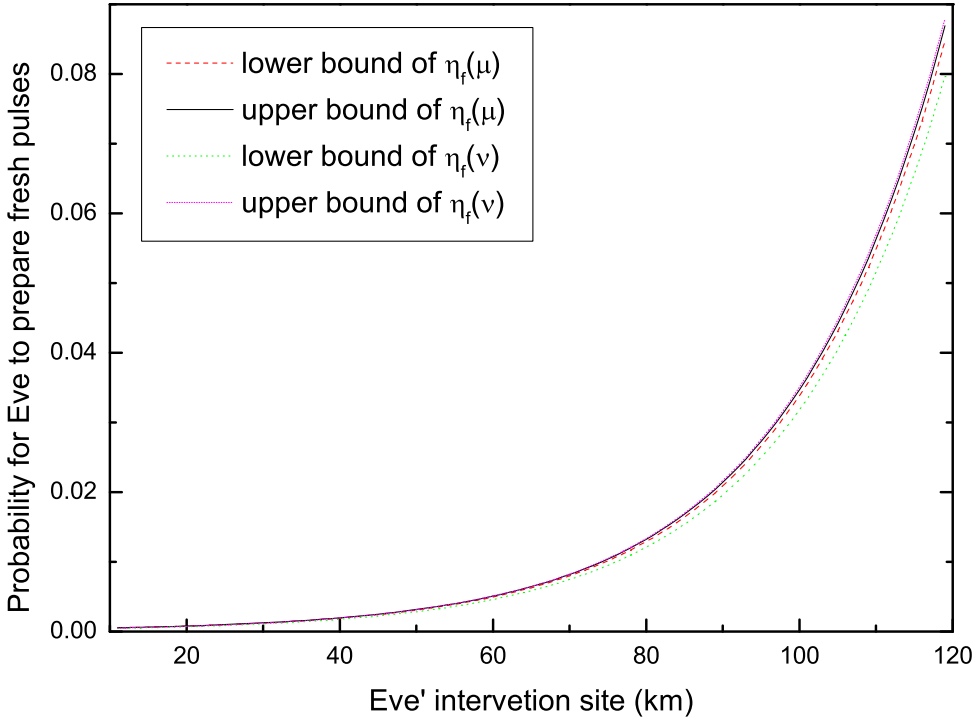}
\caption{Relationship between Eve's probability for her to prepare
fresh pulses according to her results, $\eta_{f}$ and her
intervention site, $l$. It suitable value should be chosen from the
area greater than the lower bounds and less than the upper bounds of
$\eta_{f}$ for both sources.} \label{fig:etaf}
\end{figure}

It has been proven that some commercial QKD systems may be totally
insecure under fake state
attack~\cite{Lydersen10,Yuan10,Lydersen101,Gerhardt11}, thus we hope
that decoy states is efficiency in combatting against this brutal
attack. As we have shown above, however, decoy states can not also
provide them with enough security. We have shown that Eve can launch
her fake state attack successfully. Especially, we have proven that
she can eavesdrop on the decoy state QKD without any
photon-number-resolving instrument when statistical fluctuation is
considered on Bob's results. With the presence of new technology,
especially with improvement on Bob detecting efficiency, Alice and
Bob may overcome this loophole. However, other loopholes still
unknown to people may also threaten the security of practical QKD.
And it has been claimed QKD is superior to classical cryptography as
quantum principle provide it with physically secure. In order to
avoid the mouse and cat game between legal users and eavesdropper in
QKD, new protocols should be presented to combat all these loopholes
in principle~\cite{Mayers98,Acin07,Pironio09,Lo11}. This work is
sponsored by the National Natural Science Foundation of China (Grant
No 10905028) and HASTIT.

\end{document}